\documentclass{paper}

\catcode`\|=12 

\usepackage{tikz,pst-3d,pgfplots,tikz-3dplot,xlop}
\usetikzlibrary{external}
\def\figWidth{.8\linewidth}
\def\figHeight{2cm}
\pgfplotsset{compat=1.8}

\usepackage{hyperref}
\hypersetup{
  colorlinks,
  citecolor=blue,
  linkcolor=blue,
  urlcolor=blue}

\usepackage{natbib}
\setcitestyle{authoryear,open={(},close={)}} 

\usepackage{graphicx,subcaption}%
\usepackage{amsmath,amssymb,amsfonts,physics}%
\usepackage{amsthm}%
\usepackage{mathrsfs}%
\usepackage[title]{appendix}%
\usepackage{xcolor}%

\graphicspath{{./Figures/}}
\DeclareMathAlphabet{\mathcalligra}{T1}{calligra}{m}{n}

\newcommand{\Rey}{Re}

\newcommand{\ii}{\mathrm{i}}

\newcommand{\vv}[1]{\boldsymbol{#1}}
\newcommand{\mm}[1]{\mathbf{#1}}

\newcommand{\qh}{\hat{\vv{u}}}
\newcommand{\fh}{\hat{\vv{f}}}
\newcommand{\yh}{\hat{\vv{y}}}

\newcommand{\mA}{\mm{A}}
\newcommand{\mB}{\mm{B}}
\newcommand{\mC}{\mm{C}}

\usepackage{authblk}
\begin{document}

\title{Perturbation amplification near the stagnation point of blunt bodies}


\author[1,2]{{Eduardo} {Martini}}
\author[1,2]{{Clement} {Caillaud}}
\author[1]{{Guillaume} {Lehnasch}}
\author[1]{{Peter} {Jordan}}

\author[3]{{Oliver} {Schmidt}}

\affil[1]{{Département Fluides, Thermique, Combustion}, {Institut Pprime, CNRS-Universit\`e de Poitiers-ENSMA}, {{1 Bd Marie et Pierre Curie},{ Futuroscope Chasseneuil}, {F86962}, {la Vienne}, {France}}}

\affil[2]{{Commissariat à l'Énergie Atomique}, {CESTA}, {{15 Avenue des Sablières},{Le Barp}, {331163}, {la Vienne}, {France}}}

\affil[3]{{Departamane of Mechanical and Aerospace Engineering}, {University of California San Diego}, {{ 9500 Gilman Drive},{ La Jolla}, {10587}, {California}, {USA}}}





\maketitle

\begin{abstract}
	Different transition to turbulence routes for the flow around blunt bodies are possible. Non-modal amplification of perturbations via the lift-up effect has recently been explored to explain transition near the stagnation point in axisymmetric bodies. However, only perturbations already present in the boundary layer can be amplified, and the mechanisms by which free-stream perturbations enter the boundary layer have not yet been fully explored. In this study, we present an investigation of how disturbances enter the boundary layer via the stagnation point. This linear mechanism is expected to dominate over non-linear mechanisms previously identified on the formation of boundary layer perturbations at low turbulence intensity levels. A parametric investigation is presented, revealing trends with Reynolds and Mach numbers. 
\end{abstract}

\section{Introduction}	

	Turbulent boundary layers enhance viscous drag and higher heat flux to the bodies about which they develop. Predicting where a boundary layer will transition from a laminar to a turbulent state is necessary to predict aerodynamic performance. In supersonic flight, the transition point is also a key design parameter as turbulent boundary layers exchange more heat with the surface and thus require more robust thermal shielding.
	
	Earlier work issued from local stability theory describes several paths for transition in wall-bounded flows \citep{morkovin1969many}. In scenarios where perturbations present in the free stream or due to surface roughness are small, the mechanism that leads to transition is typically modal, i.e., unstable modes in the boundary layer grow exponentially and eventually trigger transition. Another possibility is by-pass transition, which occurs for moderate perturbations. In this case, non-modal mechanisms amplify small-but-finite, disturbances. These may then reach significant amplitudes and lead to flow transition. 

	As only perturbations within the boundary layer can lead to its transition, it is necessary to understand their origin, i.e., how they penetrate to, or are generated in, the boundary layer. 
	For low Mach numbers, flow over a flat plate is typically insensitive to acoustic perturbations. The mismatch of wave number and frequency between the boundary layer structures and acoustic perturbations leads to a small coupling between them. However, \citet{goldstein1985scatteringa} has shown that points of curvature discontinuity act as a length-scale conversion mechanism, allowing acoustic waves to excite Tollmien-Schlichting (TS) waves.
	
	\cite{brandt2004transition} investigated mechanisms by which free-stream vortical perturbations can enter an incompressible boundary layer. \citeauthor*{brandt2004transition} observed that for moderate free-stream turbulence intensity, perturbations penetrate the flat-plate boundary layer mainly via non-linear interactions, becoming more relevant at larger free-stream turbulence intensities. Amplification of pre-existing boundary layer structures was dominant for lower turbulence intensities and larger integral wavelengths. The study suggests that such vortical structures may penetrate the boundary layer at the leading edge, although this mechanism was not investigated. Thus, such a linear mechanism is expected to dominate receptivity at low turbulent intensities. 
	
	For higher Mach numbers, different receptivity mechanisms may be observed. For example, in supersonic flows there are regions for which Mack waves \citep{mack1965stability} have acoustic wavelengths and can thus be directly excited by acoustic waves \citep{tam1978excitation}. 

	Several studies on this receptivity mechanism rely on numerical integration of the flow dynamics. In these, smooth surfaces, i.e., with continuous curvatures, are typically used, preventing the results from being contaminated by the length-conversion mechanism identified by \citeauthor*{goldstein1985scatteringa}.
	Leading edges shaped as super-ellipses~\citep{buter1994boundary,wanderley2001boundary,lin1992effect,shahriari2016acoustic} and parabolic bodies~\citep{haddad1998boundary} are among the examples studied. \citeauthor{haddad1998boundary} have identified that the receptivity of TS waves decreases with Reynolds number on a paraboloid body, i.e., with increasing bluntness, a result which contrasts with a more recent review on thick flat plates with curved leading edges \citep{shahriari2016acoustic}. The reason for this discrepancy is unclear, but they may be related to different flow regimes.  The study of \citeauthor{haddad1998boundary} may only be valid around the stagnation point, and, while the super-ellipse studied by \citeauthor{shahriari2016acoustic}  is formally smooth, the surrounding flow can exhibit strong gradients, whose contribution to the receptivity mechanism is unclear. 
	As a favorable pressure gradient stabilizes TS waves near the stagnation point, TS waves amplitudes are typically small around it. However, they can be amplified downstream, where the pressure gradient vanishes. Mack waves exhibit a similar behavior in supersonic flows due to the low velocities near the stagnation point.
	
	Nevertheless, there are circumstances where transition is observed close to the leading edge, e.g., in hypersonic blunt bodies \citep{paredes2017bluntbody}. For these configurations, different mechanisms have been proposed, including, on one hand, perturbation amplification in the entropy layer \citep{fedorov2004evolution,paredes2020mechanism}, which then enters the boundary layer when it swallows the entropy layer and, on the other, leading-edge modal instabilities \citep{lin1996stability,lin1997stability}. However, the former can only lead to transition after the ingestion point, which occurs further downstream than the observed transition point, and the latter are only active in the presence of a crossflow. When no crossflow is present, such modal mechanisms cannot explain these transitions, motivating the study of non-modal mechanisms in the nose region \citep{paredes2017bluntbody}. 
	
	The lift-up is a non-modal mechanism that is ubiquitous in shear flow. It consists of streamwise vortices that move fast flow from regions of higher velocities to regions of lower velocities, and vice versa, creating regions of fast and slow streamwise velocities known as streaks. Under a parallel flow assumption, linear stability theory predicts amplification factors to scale with $Re^2$ for spanwise wavelengths of the order of two boundary layer thickness in both compressible and incompressible flows \citep{andersson1999optimal,luchini2000reynoldsnumberindependent,schmid2012stability,hanifi1996transient}. Large perturbation amplification is also observed in spatially evolving boundary layers \citep{monokrousos2010global,paredes2017bluntbody,towne2022efficient,kamal2023global}. 

	The dynamics of vorticity entering the boundary layer via the leading edge have received less attention. \cite{obrist2003lineara} studied the problem using Hiemenz's self-similar flow model. However, there are virtually no studies investigating the problem for curvatures typical of a leading edge. 	
	
	In this study, we present a systematic investigation of non-modal mechanisms at the leading edge using a paraboloid body as a flow model. We systematically investigate the role of bluntness and compressibility to provide a cartography of the phenomena.
	
	The paper is structured as follows. 
	Section \ref{sec:paraboloid} presents the flow around the paraboloid body.
	Section \ref{sec:numMethod} presents the numerical methods used in the linear stability analysis.  
	Section \ref{sec:receptivity} presents aç study of the perturbation amplification mechanisms near the stagnation point using the resolvent framework.
	Section \ref{sec:compressibility} expands the results to investigate compressible flows, with Mach number up to $1.2$.  
	Section \ref{sec:conclusions} presents the final conclusions.

\section{Paraboloid body} \label{sec:paraboloid}
The body surface studied is given by 
\begin{align}
	x=(z^2+y^2)-1/4,
\end{align}
with the free-stream flow moving towards $ +z $.  The flow parameters are made non-dimensional using the freestream density, viscosity, and velocity. As the reference length, we use the curvature diameter of the leading edge.

\begin{figure}[t]
	\centering

	 \includegraphics*{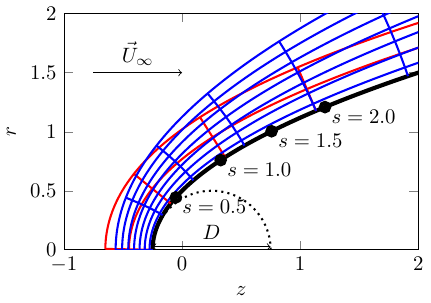}

	\caption{Illustration of the geometry and the coordinate systems used. Red and blue lines illustrate parabolic and body-fitted coordinates, respectively. The thick black line indicates the body surface. The dotted circle is tangent to the leading edge. The arrow indicates the freestream velocity. Points with a distance of multiples of $0.5$ length units from the stagnation point are shown for reference.}
	\label{fig:coordinates}
\end{figure}

Three coordinate systems are considered. Cylindrical coordinates are used in all computations, which will be detailed later. Paraboloid coordinates $(\sigma,\tau)$ are used and defined in section \ref{sec:InviscidFlow} to obtain an inviscid potential flow analytically. Finally, body-fitted coordinates $(s,d)$ are defined as the distance from the leading edge along the body surface and the distance from the body surface. They are used for visualization only. Figure \ref{fig:coordinates} illustrates the problem and coordinate systems used.

\subsection{Inviscid and incompressible flow} \label{sec:InviscidFlow}
	The potential flow around the body is obtained analytically using parabolic coordinates, $ (\sigma,\tau,\phi) $, defined as,
	\begin{align}
		 z+\ii r =& (\sigma+\ii \tau)^2, &	r =& \sqrt{(x^2 + y^2)}, &	\phi =& \tan^{-1}(y/x).  
	\end{align}
	The body surface is located at $ \sigma= \sigma_0=-1/2 $.

	The flow potential function, $ \psi $, satisfies the Laplace equation,
	\begin{align}\label{eq:laplaceParabolic}
		\nabla^2_{\sigma,\tau,\phi} \psi = \frac{1}{\sigma}\frac{d}{d\sigma}\left(\sigma \frac{d\psi}{d\sigma} \right) + \frac{1}{\tau}\frac{d}{d\tau}\left(\tau \frac{d\psi}{d\tau} \right) + \frac{d^2\psi}{d\phi^2},
	\end{align}
	with, as boundary conditions, no penetration at the body surface and the convergence to the free-stream conditions far from the body. These conditions read
	\begin{align}
		\label{eq:bc1}
		\frac{d\psi}{d\sigma}&=0 &, \text{for }  &\sigma=\sigma_0 \quad \text{and} \\
		\label{eq:bc2}
		\psi &=z=(\sigma^2-\tau^2) &, \text{for } &\sigma\to\infty .
	\end{align}

\begin{figure}
		\centering
		
		\def\figWidth{.4\linewidth}
		\def\figHeight{2cm}

		\includegraphics*{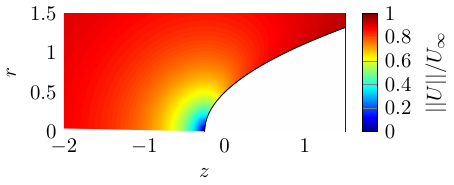}

		\caption{Potential flow around the paraboloid body.}
		\label{fig:BF_pot}
	\end{figure}

	The solution to \eqref{eq:laplaceParabolic} under \eqref{eq:bc1} and \eqref{eq:bc2} is given by 
	\begin{align}
		 \psi= \sigma^2-\tau^2 + 2\sigma_0^2 \log(\sigma).  
	\label{eq:}
	\end{align}
	The velocity field is then recovered as
	\begin{align}
		U_z = \frac{\partial \psi}{\partial x} = 
			\frac{\partial \psi}{\partial \sigma} \frac{\partial \sigma}{\partial x} + 
			\frac{\partial \psi}{\partial \sigma} \frac{\partial \tau}{\partial x},   \\
		U_r = \frac{\partial \psi}{\partial y} = 
			\frac{\partial \psi}{\partial \sigma} \frac{\partial \sigma}{\partial y} + 
			\frac{\partial \psi}{\partial \sigma} \frac{\partial \tau}{\partial y}.   
	\end{align}
	The potential flow is illustrated in figure \ref{fig:BF_pot}, where the stagnation point and subsequent flow reacceleration are observed.
	
	\subsection{Viscous incompressible flow}

	\begin{table}[]
		\centering
		\begin{tabular}{c|c|c|c|c|c|c|c|c|c}
			\multicolumn{10}{c}{\textbf{Reynolds numbers} }  \\
		$100$  & $500$ & $1\,000$& $5\,000$ & $30\,000$  &
		
		$60\,000$ & $120\,000$ & $180\,000$ & $240\,000$ &	$300\,000$ 
	\end{tabular}
	\caption{List of Reynolds numbers investigated for the incompressible case.}
	\label{tab:ReList}
	\end{table}

		A 2D axisymmetric model was used to obtain base flows for the viscous problem. No-slip boundary conditions were applied at the body surface; outflow conditions were used on the right-most edge of the domain, and the potential flow was imposed on the left-most edge as an inflow condition, reducing the domain size required to converge the solutions.
		 The problem was time-marched using the open-source code Nek5000 \citep{fischer1989parallel} until the norm of the velocity time derivative became smaller than $ 10^{-8} $. 

		This process was repeated for several Reynolds numbers between $100$ and $300\,000$, listed in table \ref{tab:ReList}. The domain size was adapted for each Reynolds number: larger domains were used at lower Reynolds numbers to minimize blockage effects, and smaller domains at larger Reynolds numbers to avoid unnecessary computational cost. The Reynolds-number range was chosen to reach values sufficiently high for an asymptotic trend to be observed, and lower values where the behavior deviates from the trend.
		
\begin{figure}
			\centering
			\def\figWidth{.4\linewidth}
			\def\figHeight{2cm}

			\includegraphics*{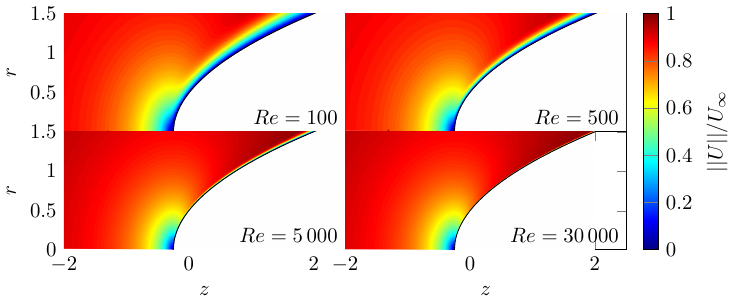}

			\caption{Velocity magnitude of the baseflow for different Reynolds numbers. For clearness, only selected cases are shown.}
			\label{fig:bfRes}
		\end{figure}

		Figure \ref{fig:bfRes} illustrates the base flows obtained. For $Re \gtrapprox 5\,000$, the flow is qualitatively similar to the potential flow shown in figure \ref{fig:BF_pot} but with the formation of a boundary layer around the body surface. The base flows are similar but with thinner boundary layers for higher Reynolds numbers.

\begin{figure}[t]
			\centering
			\def\figWidth{.8\linewidth}
			\def\figHeight{5cm}	
			\subfloat[\label{fig:bl_profiles} Wall-tangent velocity ($U_\parallel$)  profile at the wall position $x=0.3$, $s\approx0.95$.]
			{
				\includegraphics*{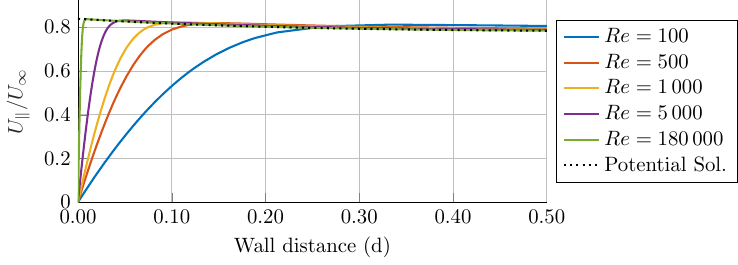}
			}

			\subfloat[\label{fig:bl_deltastar}Boundary layer displacement thickness. ]
			{
				\includegraphics*{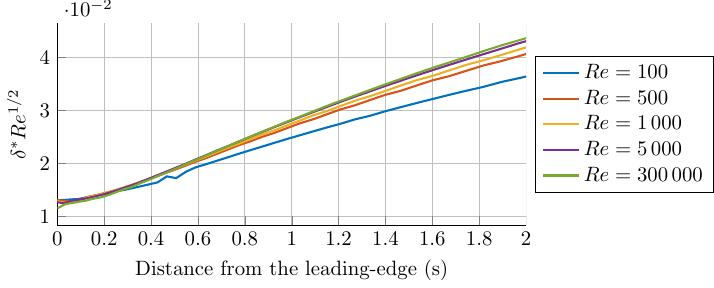}

			}
			\caption{ Boundary layer profile and thickness for the flow around the paraboloid body. }
		\end{figure}

		Figure \ref{fig:bl_profiles} shows the wall-tangential velocity profile for $s=1$, i.e., at unit distance from the stagnation point. A thicker boundary layer is formed for lower Reynolds numbers, and the flow at its edge has larger velocities than the potential flow, compensating for the loss of mass flow closer to the wall. This effect reduces with $\Rey$ and is negligible for $\Rey > 5\,000$.  As expected, the flow converges to the potential solution away from the wall for higher Reynolds numbers.

		The decreasing velocity magnitude with the potential flow's wall distance and the wall's no-slip boundary condition creates a velocity maximum at a distance $d_{max}$ from the wall. We thus define the boundary layer thickness for the problem as,
		\begin{align}
			\delta^*(s) = \int_0^{d_{max}} \left(\frac{U_{\parallel,\text{pot}}(s,d)-U_{\parallel}(s,d)}{U_{\parallel,\text{pot}}(s,0)}\right) \dd d.
			\label{eq:deltas}
		\end{align}
		where $U_{\parallel}$ indicates velocity component parallel to the wall, and $U_{\parallel, \text{pot}}$ corresponds to the potential solution.  Note that $U_{\parallel,pot}(s,0)=1$ for a flat plate, and \eqref{eq:deltas} recovers the classical definition of the displacement thickness.

		Figure \ref{fig:bl_deltastar} shows the evolution of the boundary layer profile. Near the stagnation point, the boundary layer is formed with a thickness that scales with $\Rey^{-1/2}$, as predicted when the region is modeled by a Homann flow \citep{homann1936einfluss,weidman2012nonaxisymmetric}. Downstream, the boundary layer evolution reaches the asymptotic limit where it scales with $Re^{-1/2}$ for $\Rey\gtrapprox5\,000$.

		\section{Numerical method}	\label{sec:numMethod}

		We will perform resolvent analysis on a 2D axisymmetric flow to study the perturbation amplification.  The numerical method is derived from that used by \cite{schmidt2018spectral} and \cite{schmidt2011linear,schmidt2014viscid}, with the code adapted to use curvilinear coordinates. 
		We refer readers to the original document for details of the code, describing next only its main features.
		
		\subsection{Resolvent analysis}
		The discretized linearized Navier-Stokes equations written in input-output form in the frequency domain 
		\begin{align}
			\begin{aligned}
			(-\ii \omega- \mA) \qh (\omega) &= \mB \fh (\omega),   \\
			 \yh (\omega) &= \mC \qh (\omega),
			\end{aligned}\label{eq:linNS}
		\end{align}
		where $ \mA $ is the discretized representation of linearized Navier-Stokes operator around a fixed point, $ \mB $ and $ \mC $ are matrices introduced to limit input (forcing) and output (responses) spaces. E.g., these can be used to limit the forcing and response to regions near the boundary layer, avoiding the contamination of the results by free-stream modes \citep{nogueira2020resolvent}. The forcing and response vectors are, respectivelly, $\fh = [\hat f_\rho,\hat f_{u_z},\hat f_{u_r},\hat f_{u_\theta},\hat f_T]^T$, and $\qh = [\hat \rho,\hat u_z,\hat u_r,\hat u_\theta,\hat T]^T$ .

		A direct relation between input and the output is obtained as
		\begin{equation}\label{key}
			\yh(\omega)=\mm{R}(\omega) \fh(\omega), 
		\end{equation}
		where 
		\begin{equation}\label{eq:LinSys}
		 \mm{R}(\omega )= \mC 	(-\ii \omega- \mA) ^{-1}  \mB ,
		\end{equation}
		is the resolvent operator \citep{schmid2012stability}.
		
		A forcing term is said to be optimal if it maximizes the Rayleight ratio,
		\begin{equation}\label{key}
			\sigma^2 (\omega)
				= \dfrac{ \left\lVert  \yh(\omega)  \right\lVert _{\mm{W}_y} }  {\left\lVert\fh(\omega) \right\lVert _{\mm{W}_f}}  
				= \dfrac{ \left\lVert \mm{R}(\omega) \fh(\omega)  \right\lVert_{\mm{W}_y} } {\left\lVert\fh(\omega) \right\lVert _{\mm{W}_f}},
		\end{equation}
		and can be obtained from a singular value decomposition (SVD) of the weighted resolvent operator $ \mm{R}'(\omega) = \mm{W}_y^{-1/2} \mm{R}(\omega) \mm{W}_f^{1/2} $, where $ \mm{W}_{y/f} $ are the weight matrices for the response and forcing, respectively. In this study, the weight matrices include integration quadrature weight and an energy norm for compressible flows. Two different energy norms are used, a kinetic,
			\begin{align}
			  \lVert \yh \rVert^2 = \int \overline{\rho} \left(  |\hat{u}_z|^2 +|\hat{u}_\theta|^2 +|\hat{u}_r|^2  \right) \dd V 
			\label{eq:energyNorm_kin}
			\end{align}
		and a compressible \citep{chu1965energy,hanifi1996transient},
		\begin{align}
			\lVert \yh \rVert^2 = \int \left(
			\frac{\overline{T}}{\overline{\rho}\gamma M^2}|\hat{\rho}|^2 
			+\overline{\rho} |\hat{u}_z|^2 
			+\overline{\rho} |\hat{u}_\theta|^2 
			+\overline{\rho} |\hat{u}_r|^2 
			\frac{\overline{\rho}}{\overline{T}\gamma(\gamma-1) M^2}|\hat{T}|^2 
			\right) dV.
			\label{eq:energyNorm_chu}
		\end{align}
		norms, where the overline variables refer to the baseflow fields, $\gamma$ is the adiabatic coefficient, and $M$ is the Mach number.
		
		Different strategies to compute the SVD can be used.  
		For problems with only one inhomogeneous direction, $ \mm{R}(\omega) $ is a small matrix, and standard tools are effective in obtaining the decomposition. On the other hand, for three-dimensional problems, methods based on time-marching schemes are typically used, which can be based on matrix-free approaches~ \citep{monokrousos2010global,gomez2016reduced,martini2021efficient,farghadan2021randomized}. For two-dimensional problems where one direction is homogeneous, one of the most effective approaches is the use of sparse matrices to represent $ \mA $. In this approach, the effect of the resolvent operator on a vector is obtained by solving the linear system \eqref{eq:linNS}, which can be effectively obtained using a lower-upper (LU) decomposition of its left-hand side. The latter approach is employed in this work.
		
		\subsection{Curvilinear Coordinates}
\begin{figure}
			\centering
			\includegraphics[scale=0.4]{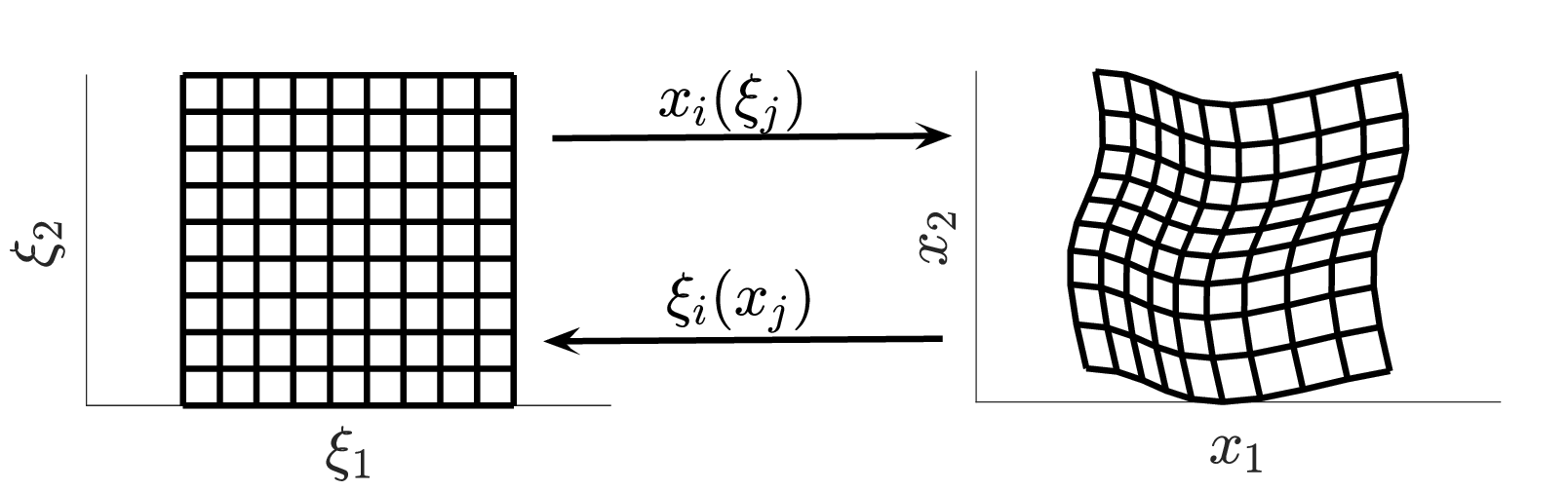}
			\caption{Illustration of mapping between the computational ($ \xi_i $) and physical ($ x_i $) meshes.}
			\label{fig:meshdeformation}
		\end{figure}

		In order to construct the linearized operator, derivative matrices for the domain are required. These matrices were constructed by mapping a square domain (described with coordinates $\xi_i$), for which derivatives are computed using standard finite differences, into the desired domain (described with coordinates $x_i$), as illustrated in figure \ref{fig:meshdeformation}.  Centered and uncentered fourth-order finite-difference schemes were used on the bulk of the domain and at the domain boundaries, respectively.

		In general, the mapping between $ \xi_i $ and $ x_i $ is given implicitly: for a given node in the original mesh, $ x_i(\xi_j) $ corresponds to the coordinate of the same node on the mapped domain. The inverse function, $ \xi_i(x_j) $, is likewise available.  Henceforth, we will refer to $ x_i(\xi_j) $ simply as mapping, and to $ \xi_i(x_j) $ as inverse mapping. 
		
		Using the chain rule, the derivative of a function $ g $ is written as
		\begin{equation}\label{eq:chainRule}
			\partial_{x_i} g = {J}_{ij}	\partial_{\xi_j} g,
		\end{equation}
		where $ J_{ij} = \frac{\partial \xi_i}{\partial x_j}$ are $(i,j)$ entries of $\mm{J}$, the Jacobian of the mapping function. As only differentiation matrices w.r.t. $\xi_i$ are a priori available, we construct first $J_{ij}^{-1} = \frac{\partial x_j}{\partial \xi_i} $. We then obtain $\mm{J}$ by inverting $\mm{J}^{-1}$.
		
		Eq. \eqref{eq:chainRule} can be iterated to obtain second-order derivatives. However, this increases the effective stencil of the method, which increases the cost of solving the resulting linear systems.  To avoid larger stencils, we  describe the second derivative as
		\begin{align}
			\begin{aligned}
		\partial_{x_i}\partial_{x_j} g  
			&= \partial_{x_i} \left( {{J}}_{jq} 	\partial_{\xi_p} g \right)  \\
			& =  \left(  \partial_{x_i}  {{J}}_{jp} \right) 	\partial_{\xi_p} g   + 
				 {{J}}^{-1}_{jp}  \left(  \partial_{x_i}  	\partial_{\xi_p} g  \right) \\
			& =  \left(  \partial_{x_i}  {{J}}_{jp} \right) 	\partial_{\xi_p} g   + 
					{{J}}^{-1}_{jp}   {{J}}_{iq} 	\partial_{\xi_q} 	\partial_{\xi_j} g,
					\label{eq:chainRule2}
			\end{aligned}
		\end{align}
		computing  $ \left(  \partial_{x_i}  {{J}}_{ij} \right)  $ via the matrix identity 
		\begin{equation}\label{key}
		   \partial_{x_i}  {{\mm J}}  =  {\mm J} \left( \partial_{x_i}  {\mm J}^{-1}  \right)  {\mm J}.
		\end{equation}
		
		Finally, the term $ \partial_{x_i}  {\mm J}^{-1}  $, reads
		\begin{align}\label{key}
			\partial_{x_i} J_{lm}^{-1}  & =	J_{ji}^{-1} \partial_{\xi_i} J_{lm}^{-1} = J^{-1}_{ij} \dfrac {\partial^2 x_l} {\partial \xi_i \partial \xi_m}. 
		\end{align}
		
		With this approach, the stencil size in the physical domain ($x_i$) has the same size as that of the computational domain ($\xi_i$).

	\section{Receptivity} \label{sec:receptivity}
		We start by investigating the receptivity of the $Re=30\,000$ baseflow. A Cartesian grid with $100\times 100$ points was mapped around the body, with nodes clustered near the surface to resolve the boundary layer. The convergence of all the results presented was checked, and variations less than $1\%$ when using a finer mesh were observed.

		As will be shown next, the spatial support of the forcing/response modes can extend far up/downstream. Fully capturing these structures requires a large domain, making the analyses too costly. Since the study aims to investigate how free-stream disturbances penetrate the boundary layer, the downstream domain was limited to $s\lessapprox 2$. The domain upstream was limited to $z\gtrapprox-6$. Down/upstream of these limits, a sponge zone is used to dampen perturbations before the boundary of the computation domain.

\begin{figure}
			\centering
			\def\figWidth{.8\linewidth}
			\def\figHeight{5cm}
			\includegraphics*{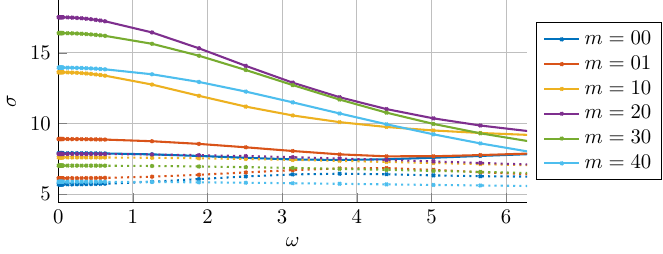}

			\caption{Gain variation with frequency for the $Re=30\,000$ flow. Leading/suboptimal gains in solid/dashed lines for different azimuthal numbers $m$.}
			\label{fig:gains_freq}
		\end{figure}

		Figure \ref{fig:gains_freq} shows the frequency dependency of the leading gains for different azimuthal wavenumbers. The largest gain is found for $\omega=0$, first evidence that the lift-up mechanism is dominant around the leading edge. We henceforth concentrate the study on gains associated with zero-frequency disturbances.

		\begin{figure}
			\centering
			\subfloat[ \label{fig:gains_opt} Leading gains]{
				\includegraphics*{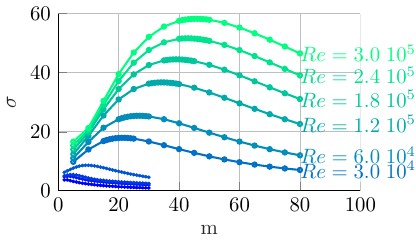}
				}
			\subfloat[ \label{fig:gains_subopt} Gain separation]{
				\includegraphics*{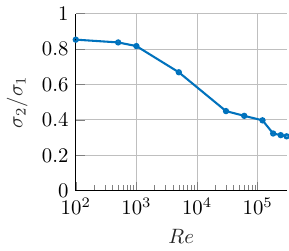}
				}
			\caption{Leading gains (a) and suboptimal gains for the dominant azimuthal number (b). The unlabelled lines in (a) correspond to $Re=100,500,1\,000,$ and $5\,000$, from the bottom up.}
			\label{fig:gains}
		\end{figure}
		
		We next perform a systematic parametric investigation, computing gains at $\omega=0$ for different Reynolds and azimuthal numbers. The numerical grid was adapted for each scenario as the boundary layer thickness varied considerably between the different cases. The Cartesian grid size between $75\times75$ and $ 100\times 175$ gridpoints were used, with the last number corresponding to the wall-normal direction near the body surface, to discretize the boundary layer.
		The results are presented in Figure \ref{fig:gains_opt}, which shows that both the maximum gain and the corresponding azimuthal wavenumber increase with $Re$. Figure \ref{fig:gains_subopt} shows that the difference between the leading and the suboptimal gains tends to also increase with $\Rey$, which points towards an increasing selectivity of disturbances.

		\begin{figure}[t]
			\centering
			\def\figFolder{Figures/Re5kmodes}
			\def\figWidth{.4\linewidth}
			\def\figHeight{1cm}

			\subfloat[][]{
				\includegraphics*{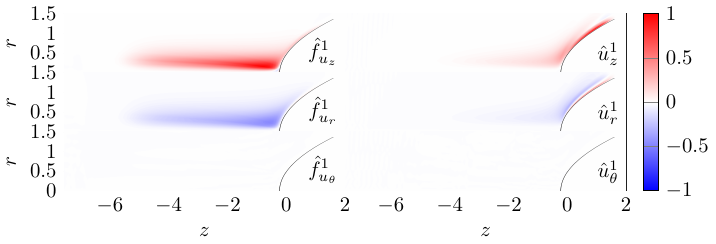}
			}
			
			\subfloat[][]{
			\includegraphics*{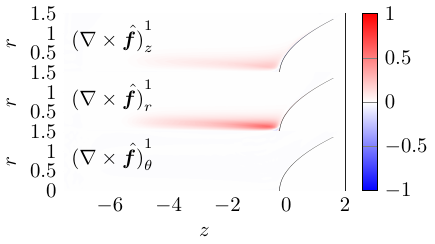}
			}

			\caption{Leading forcing and response modes (a) and the curl of the forcing mode (b) for  $ \Rey=5\,000 $ and $m=9$. The modes are normalized by their largest absolute value. Color scales indicate the real part of the modes.}
			\label{fig:bfre5000}
		\end{figure}

		\begin{figure}
			\centering
			\def\figFolder{Figures/Re5kmodes}
			\def\figWidth{.4\linewidth}
			\def\figHeight{1cm}

			\includegraphics*{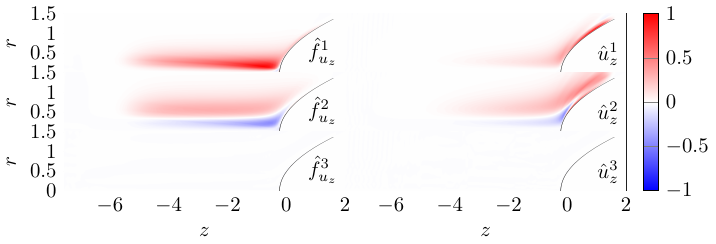}

			\caption{Leading and suboptimal forcing and response modes. Results for $ \Rey=5\,000 $ and $m=9$. The modes are normalized by their largest absolute value. Color scales indicate the real part of the modes. }
			\label{fig:bfre5000_sub}
		\end{figure}

		\begin{figure}[t]
			\centering
			\def\figWidth{.4\linewidth}
			\def\figHeight{1cm}
			\includegraphics*{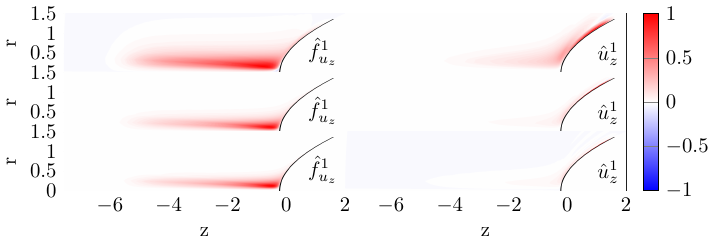}

			\caption{Spatial support of the leading forcing and response modes.  From top to bottom, results for $Re=5\,000$, $30\,000$, and $60\,000$, with azimuthal wave number corresponding to the highest gain for each $Re$. The modes are normalized by their largest absolute value. Color scales indicate the real part of the modes.}
			\label{fig:bf_ReComparison}
		\end{figure}

		To illustrate the forcing and response modes, Figure \ref{fig:bfre5000} shows their support for the highest gain observed for $\Rey=5\,000$ and $m=9$. Although this case has a moderate gain, the relatively thick boundary layer facilitates visualization of the flow features. The modes for higher Reynolds numbers are qualitatively similar. Both forcing and response modes are concentrated in the $z$ and $\theta$ components. The forcing mode excites vorticity perturbations upstream of the leading edge. Over the surface, the response mode is mostly dominated by the streamwise velocity component. These observations suggest that the dominant mechanism at play consists of free-stream streamwise vortices penetrating the boundary layer at the stagnation point and creating streamwise streaks via the lift-up mechanism. The $\rho$ component of the vorticity being active in creating the streaks near the stagnation point; and the $z$ component being active further downstream, where the flow becomes more aligned with the $z$ axis.  
		 The spatial support of the forcing modes extends upstream with the Reynolds numbers, and thus, resolving it requires large domains. We limit the upstream domain, as it captures the relevant physical mechanism.

		Figure \ref{fig:bfre5000_sub} compares the support for the optimal and suboptimal force and response modes. The suboptimal modes exhibit nodes in the radial direction for both the forcing and response modes, similar to the streak force and response modes in jets \citep{wang2021effecta}. The sub-optimal modes for all cases studied here presented a similar structure.

\begin{figure}[t]
			\centering
			\def\figWidth{.9\linewidth}
			\def\figHeight{3cm}

			\includegraphics*{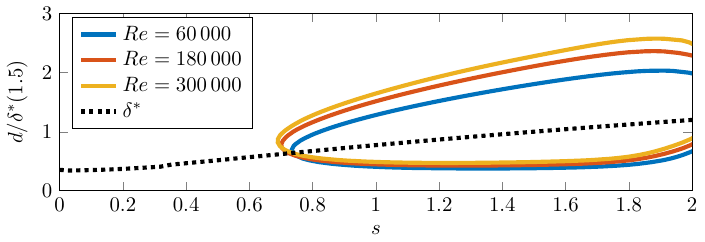}

			\caption{Isocontours of half of the leading response mode peak value for different Reynolds numbers. The displacement thickness for $\Rey=300\,000$ is shown for reference.  }
			\label{fig:modes_comparisons}
		\end{figure}

		Evolution of the modes with Reynolds number is shown in figure \ref{fig:bf_ReComparison}. The forcing modes move close to the axis of symmetry for higher $\Rey$, which can be interpreted as the region for which the streamlines converge to, or close to, the boundary layer. The response modes are increasingly concentrated in the boundary layer and move upstream and closer to the body surface with $\Rey$.
		Figure \ref{fig:modes_comparisons} shows the contour of the response modes scaled by the boundary layer thickness using the wall-fitted coordinates $(s,d)$, previously shown in figure \ref{fig:coordinates}. The modes are distributed around the boundary layer displacement thickness. 		A slight wall-normal displacement of the modes is observed for the highest $\Rey$ numbers.
		Whether this trend persists for higher Reynolds numbers or its implication is unclear.

\begin{figure}[t]
			\centering

			\includegraphics*{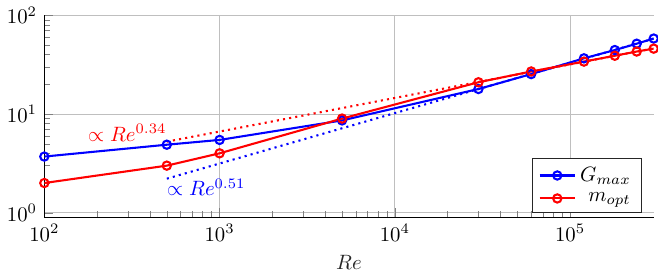}

			\caption{Reynolds number scaling of the maximum gain and the corresponding azimuthal number. Dotted lines show a linear fit for $Re\ge3\times10^4$. }
			\label{fig:scallings}
		\end{figure}

		Figure \ref{fig:scallings} shows the trends associated with the optimal gains, $\sigma_{opt}$, and the corresponding azimuthal wavenumbers, $m_{opt}$, with $\Rey$. For $\Rey\ge 30\,000$, $\sigma_{opt}$ follows a $\Rey^{1/2}$ scaling, while $m_{opt}$ scales with $\Rey^{0.34}$. 
		These scalings contrast with those of a flat plate in two respect. The gains scale with the boundary layer thickness ($\Rey_{\delta^*}=\Rey^{1/2}$), instead of its square, as found for flat plates. This difference is possibly due to the finite domain used here: note that the flat-plate $\Rey$ scaling is obtained for locally parallel approximations, which assumes infinite domains. Another trend is the increase in the spanwise wavelength in terms of the boundary layer thickness, with $m \propto \Rey^{0.34} =\Rey_{\delta^*}^{0.68}$. We speculate that this difference may be due to the axisymmetric geometry: a given value of $m$ implies a different ``spanwise'' wavelength ($2\pi r/m$) for different positions at the body. Different azimuthal numbers are optimally amplified locally at each position as the boundary layer thickness increases with the distance to the stagnation point. A trade-off between different amplification rates at different positions may be responsible for the observed trend. 
	
		\section{Compressibility effects}\label{sec:compressibility}
\begin{figure}
			\def\figWidth{.4\linewidth}
			\def\figHeight{2cm}

			\centering
			\includegraphics*{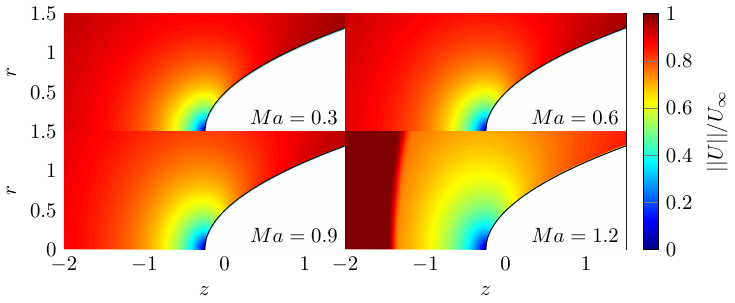}

			\caption{Illustration of the velocity profiles for different Mach numbers for $Re=3\times 10^4$.}
			\label{fig:BaseFlowsComp}
		\end{figure}

\begin{figure}
			\def\figWidth{.8\linewidth}
			\def\figHeight{5cm}

			\centering
				\includegraphics*{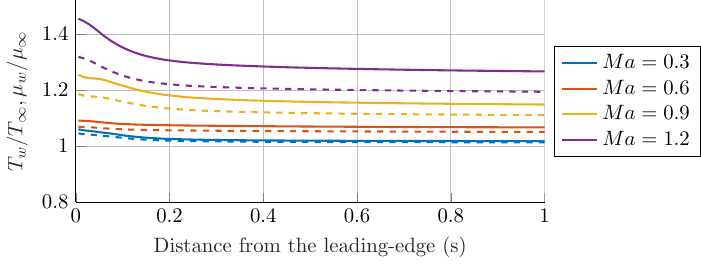}

			\caption{Viscosity (dashed lines) and temperature (solid lines) along the paraboloid wall.}
			\label{fig:BaseFlowsComp_wallProp}
		\end{figure}

		Having explored receptivity of the incompressible paraboloid flow, we now assess the effect of compressibility for increasing Mach numbers.
		We investigate Mach numbers $0.3$, $0.6$, $0.9$, and $1.2$, and fix the  Reynolds number to the value of $30\,000$, for which the asymptotic trend observed in Figure \ref{fig:scallings} is reached. Baseflows are obtained using the code SU2 \citep{economon2016su2}. Freestream values and adiabatic walls are used as boundary conditions.  Baseflows were converged until the residuals were smaller than $10^{-8}$. The resulting baseflows are illustrated in Figure \ref{fig:BaseFlowsComp}. Figure \ref{fig:BaseFlowsComp_wallProp} shows the temperature and viscosity profile along the body surface. Higher temperature and viscosity are observed near the stagnation point.

		We smoothed the shock discontinuities for the stability analysis of the supersonic flow cases, as shown in figure \ref{fig:BaseFlowsComp}, and we used a finer mesh with $200 \times 200$ grid points. We considered two cases, one using a kinetic energy norm \eqref{eq:energyNorm_kin}, i.e., restricting the force- and response-mode norm to the velocity components, and another considering temperature and density variations, using a compressible energy norm \eqref{eq:energyNorm_chu}. While it can be argued to be more generic, it has been argued that velocity perturbations are more likely to trigger flow transition. We thus choose to study both.

\begin{figure}
			\centering
			\def\figWidth{.8\linewidth}
			\def\figHeight{5cm}
			\includegraphics*{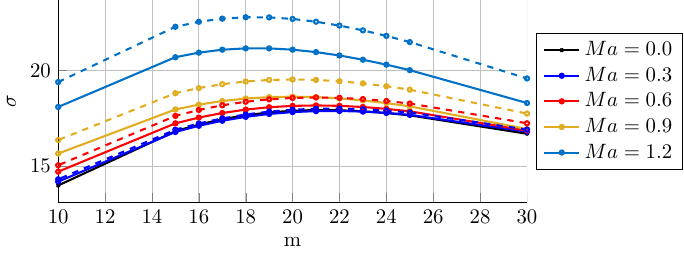}

			\caption{Gains for the leading zero frequency mode for different Mach numbers. Solid lines indicate results where forcing and responses are restricted to the momentum equations and velocity components. Dashed lines consider temperature and density fluctuations.}
			\label{fig:gains_comp}
		\end{figure}

\begin{figure}
			\centering
			\def\figHeight{3cm}

			\includegraphics*{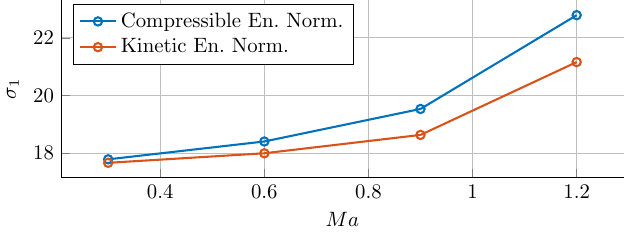}

			\caption{Gains trend with Mach number.}
			\label{fig:gains_trend_comp}
		\end{figure}

		Figure \ref{fig:gains_comp} shows the gains associated with flows with increasing Mach numbers. Results are similar to those of the incompressible flow, with a monotonic increase in value with Mach number, which is consistent with the results presented by \citet{tumin2001spatial}. The reduction of the optimal spanwise wavenumber shown in their study is also observed in figure \ref{fig:gains_comp}. The trend of the peak gain for each condition is summarized in figure \ref{fig:gains_trend_comp}.

\begin{figure}
			\centering
			\def\figFolder{Modes_cartesian}
			\def\figWidth{.4\linewidth}
			\def\figHeight{1cm}

			\includegraphics*{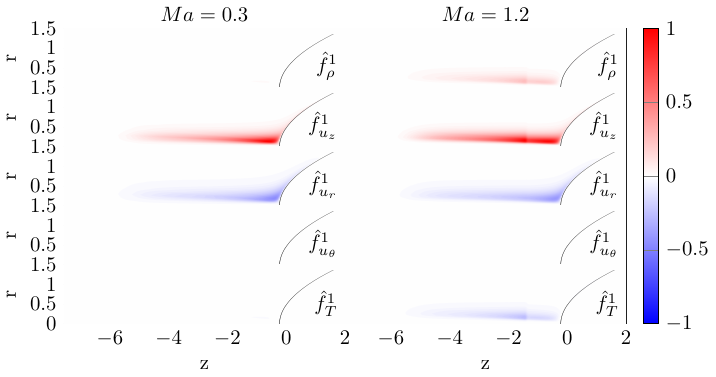}

			\includegraphics*{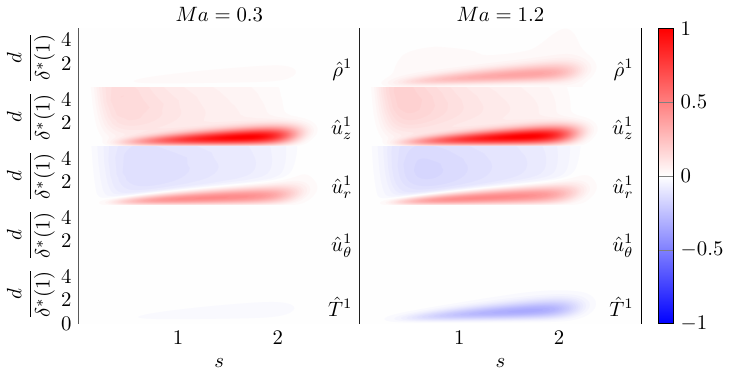}

			\caption{Forcing and response modes for $m=21$ modes for the compressible flow cases. Color scales indicate the real part of the modes.}
			\label{fig:compressibleModes}
		\end{figure}

		The support of the optimal forcing and response modes are shown in figure \ref{fig:compressibleModes}. The support of the forcing terms of the momentum equations remains qualitatively similar, and the shock, located around $z\approx2$, does not seem to affect these terms. Extra forcing terms for the continuity and energy equations appear, showing a transition for the regions before and after the shock.  Again, the flow responses are qualitatively similar, but with density and temperature fluctuations becoming more pronounced for higher Mach numbers.

\section{Conclusions} \label{sec:conclusions}	 

	We have presented a parametric investigation of non-modal mechanisms at the leading edge of blunt bodies.  Reynolds- and Mach-numbers trends associated with the perturbations optimally amplified by the flow were identified. 
	
	Optimal perturbations were obtained using resolvent analysis. To study a paraboloid body, a curvilinear grid formulation was implemented in a linear stability code. The baseflow used for analysis was obtained by time marching the non-linear flow equations.
	
	Close to the stagnation point, maximum perturbation amplification occurs for zero-frequency disturbances. The optimal modes indicate a receptivity mechanism associated with free-stream vorticity that enters the boundary layer in the vicinity of the stagnation point and then uses the lift-up effect to create streaky structures at the body surface.
	  
	In the incompressible regime, a parametric investigation reveals that for increasing Reynolds number, i.e., increasing bluntness, the maximum amplification factor increases with $Re^{1/2}$. The result contrasts with the reduction of receptivity of Tollmien-Schlichting waves with bluntness previously reported for the same geometry. Increases in Mach number also lead to larger perturbation amplification. These trends reinforce the importance of non-modal mechanisms for blunt bodies at high Reynolds numbers in the incompressible and compressible limits.

	The results complement previous receptivity studies. The trends presented are consistent with the switch from modal to non-modal transition mechanisms in hypersonic flows around blunt bodies \citep{paredes2017bluntbody}. In the low Mach regime,  the decreasing receptivity of Tollmien-Schlichting waves with bluntness \citep{haddad1998boundary} points towards a similar phenomenon in incompressible flow: as bluntness scales, the role of non-modal mechanisms tends to overcome that of modal mechanisms, potentially leading to a transition to turbulent close to the stagnation point.
	
	Finally, the linear mechanism by which free-stream disturbances penetrate the boundary layer complements previously studied transition routes for flow over a flat plate. While for moderate free-stream disturbances in the incompressible limit, non-linear receptivity is the dominant mechanism by which perturbations enter the boundary layer \cite{brandt2004transition}, at lower turbulence intensities, a linear receptivity mechanism tends to dominate. We have thus studied one such mechanism:  the ingestion of vortical disturbances at the stagnation point.

	This work also complements previous studies of the blunt body paradox by exploring perturbation amplification trends near the stagnation point. The paradox comes from observations that, in hypersonic flows, the transition points move downstream with increasing bluntness up to a critical bluntness, after which it quickly moves upstream. The trends of the perturbation amplification with Reynolds number presented in this work is consistent with investigations indicating that the transition reversal is due to a non-modal mechanism, and can be used as a departure point for further studies of the phenomena.

	\section*{Acknowledgments}
	This work is part of the project TRANSITION supported by Région Nouvelle-Aquitaine under grant 2018-1R10220. This work has been granted access to the HPC resources of IDRIS under the allocations A0092A10868 and A0112A10868 made by GENCI (Grand Equipement National de Calcul Scientifique).

\bibliographystyle{abbrvnat}

\bibliography{References.bib}

\end{document}